\documentclass{article} 
\usepackage{mrs2005,epsfig}
\begin{document} 
\title{The Millennium Galaxy Catalogue: The nearby supermassive black hole mass function}

\author{Simon P. Driver, Alister W. Graham, Paul D. Allen}
\affil{RSAA, Mount Stromlo Observatory, AUSTRALIA} 
\author{Jochen Liske}
\affil{ESO, Garching, GERMANY}

\begin{abstract} 
We highlight the correlation between a galaxy's supermassive black
hole mass and the S\'ersic-index of the host spheroid or bulge
component. From our bulge-disk decompositions of 10 095 galaxies,
drawn from the {\it Millennium Galaxy Catalogue}, we construct the
local ($ z < 0.18$) mass function of supermassive black holes. We
compare our results to those of McLure \& Dunlop (2004) and conclude
that the mass density of supermassive black holes may be marginally
higher than previously supposed. This increase is predominantly due to
the inclusion of low mass and later-type bulges. More details will be
presented in a forthcoming paper.
\end{abstract} 
 
\vspace{-1.0cm}

\section{Introduction} 
Supermassive black holes ($10^{9}$---$10^{6}$M$_{\odot}$) have now
been identified through core velocity measurements for $\sim 40$
nearby systems, see for example the compendium of Ferrarese \& Ford
\cite{ferrarese}. The formation mechanism of these super massive black
holes remains uncertain. Silk \& Rees \cite{silkrees} advocate
formation directly through initial monolithic collapse whereas Croton
et al. \cite{croton} advocate a gradual build-up through successive
mergers. It is also unclear whether central black holes occur within
all spheroidal, bulge and possibly nucleated systems.  In particular,
their existence (or the existence of intermediate-massive black
holes), in dwarf systems is currently being pursued with no definitive
result as yet, although see \cite{valluri}. As well as their recent
detection we are also now aware of the tight correlation between the
black hole mass and the global velocity dispersion ($\sigma$) of the
bulge or spheroid which they inhabit
\cite{ferrarese00},\cite{gebhardt}. From the $L-\sigma$ relation
\cite{fj} it also follows that there is a correlation between the
black hole mass and the bulge-luminosity, albeit possibly non-linear
\cite{Matkovic}.

Here we highlight a third correlation between the concentration of a
galaxy bulge or spheroid and the supermassive black hole mass
\cite{graham01}. This correlation is to be expected given the known
relation between the luminosity, velocity dispersion and
concentrations of nearby spheroid and bulge systems \cite{trujillo},
\cite{graham01b}. Which of the three relations ($M_{BH}-\sigma$,
$M_{BH}-L$ or $M_{BH}-n$) is the more fundamental is not yet clear. In
practical terms the S\'ersic index is arguably the easier to measure
as it a purely photometric quantity requiring neither kinematic
observations ($\sigma$) nor absolute photometric calibration ($L$).

\section{Concentration and the S\'ersic index}
The concentration of a galaxy's light is typically given as the ratio
of two radii. For example the SDSS adopt the radius which encompasses
90\% of the total flux divided by the radius which encompasses 50\% of
the total flux, i.e.,

~

$C_{(90/50)} = \frac{R_{90}}{R_{50}}$.

~

\noindent
Alternate definitions are in common practice which use different
ratios (e.g., \cite{conselice} who use $C_{80/20}$ for example). The
concentration index can be related analytically to the S\'ersic index
(see recent review by \cite{gd05}) of the projected light-profile
\cite{graham01b}, where the S\'ersic intensity model (\cite{sersic63},
\cite{sersic68}), is given by:

~

\begin{figure}[h]
\begin{center}
\epsfig{figure=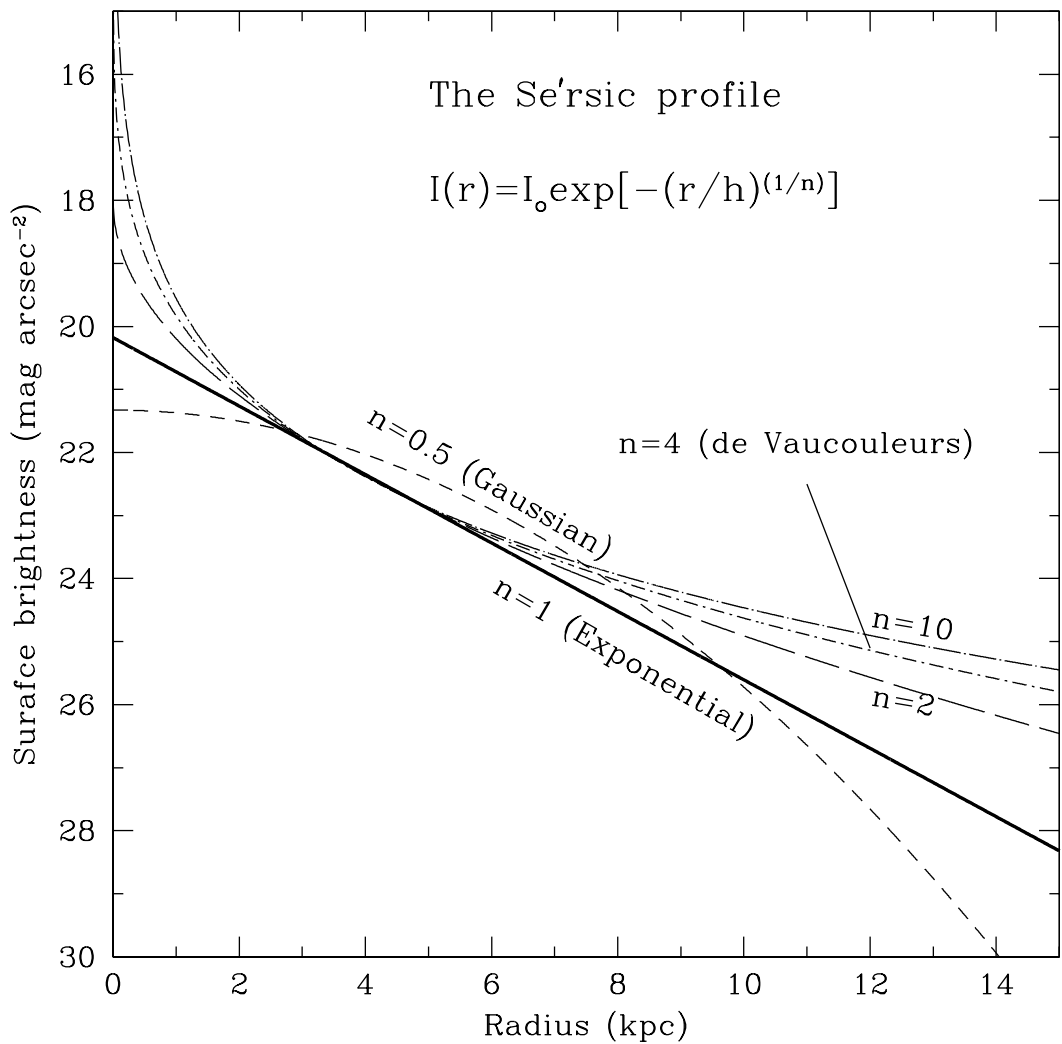,width=15.0cm}  
\end{center}
\caption{Illustration of the projected S\'ersic profile with index
varying from 0.5 to 10 as indicated. All profiles have the same
half-light radius and surface brightness at the half-light radius but
differing magnitudes. \label{fig:profiles}}
\end{figure}

$I(r)=I_o \exp[-(\frac{r}{h})^{(1/n)}]$,

~

\noindent
where I(r) is the intensity at radius, r, $I_o$ the central intensity,
$h$ the scale-length, and $n$ the S\'ersic index or profile shape
parameter. Note that $n=1$ refers to the traditional exponential
profile \cite{kcf70} and $n=4$ the traditional de Vaucouleurs profile
\cite{dv}. A value of $n=0.5$ represents a Gaussian-shaped profile
(see Fig.~\ref{fig:profiles}).

\section{The SMBH-n correlation}
\begin{figure}[h]
\begin{center}
\epsfig{figure=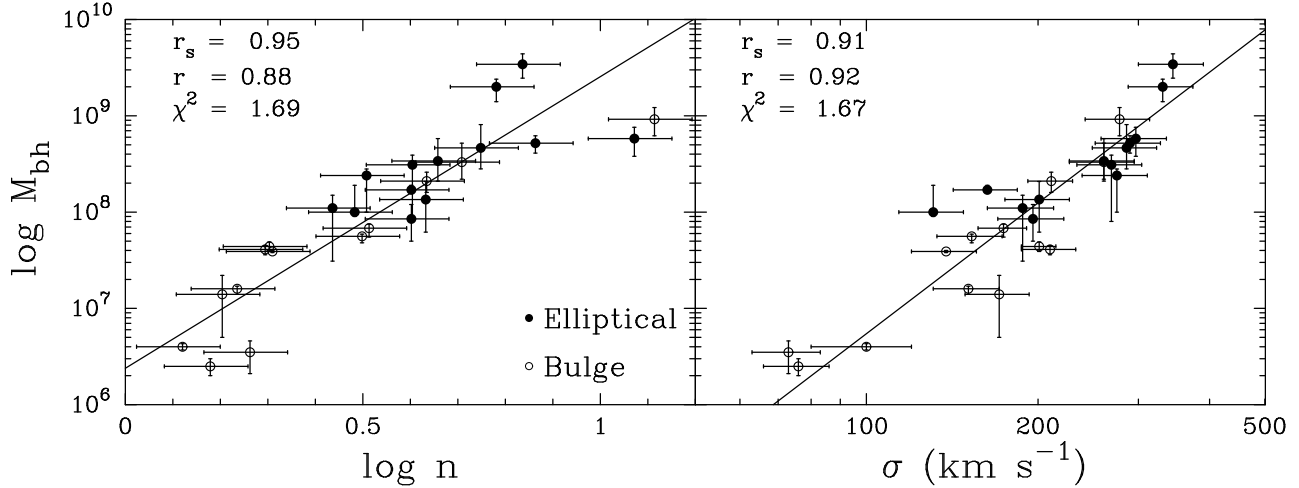,angle=-90,width=\textwidth}  
\end{center}
\caption{({\it left}) supermassive black hole mass versus S\'ersic
index of the bulge component or spheroid, and ({\it right})
supermassive black hole mass versus bulge or spheroid velocity
dispersion (see section 3). Nonlinear fits are explored in Graham, et
al (2005, in prep.) \label{fig:smbh-n}}
\end{figure}

Fig.~\ref{fig:smbh-n}({\it left}) shows the $M_{BH}-n$ relation while
Fig.~\ref{fig:smbh-n}({\it right}) shows the $M_{BH}-\sigma$ relation
for the same systems.  The linear fits are by ordinary least squares
regression assuming a 20\% uncertainty on n.  Both distributions show
a strong correlation with comparable (linear) Pearson, $r$, and
(non-linear) Spearman, $r_s$, rank-order coefficients. Work is
underway to add the remaining dozen or so systems to this plot for
which reliable black hole masses and distances are accurately
known. Two further issues are worth mentioning. (1) The S\'ersic-index
measurements are predominantly made in the $R$ filter. While this is
less susceptible to dust attenuation/distortion (and particularly for
the spiral galaxy bulges), near-IR profiles would be more optimal as
the S\'ersic index can be wavelength dependent (because of dust which
is more centrally concentrated than the stars). (2) The sample is not
entirely random but reflects those systems sufficiently nearby such
that black hole mass constraints can be obtained. As we live in a
specific region of the Universe (a loose group) this may bias our
sample in some unforeseeable way. However, as the loose group
environment appears to be the most common \cite{eke} any bias is
likely to be small and subtle. The best linear fit to our data is:
\begin{equation}
\log M_{BH} = (3.03 \pm 0.48) \log (n/3) +7.82 (\pm 0.07).
\end{equation}
This relation can be used to construct the SMBH mass function for any
sample of galaxies for which the S\'ersic-index has been measured. As
measuring the S\'ersic-index is much easier than the (bulge) velocity
dispersion, which can be particularly problematic to measure for
late-type systems due to disk/bulge contributions, one might expect
the mass function derived via this method to be more robust.

\section{The MGC and bulge-disk decomposition}
The {\it Millennium Galaxy Catalogue} (MGC, \cite{mgc1}, \cite{mgc4})
has been discussed in two earlier talks at this meting (see Allen and
Liske these proceedings). It constitutes a 37.5 sq degree imaging
survey along the Northern Spring equatorial strip with redshift
information for $96$\% of the galaxies for $B < 20$ mag. The sample
has a median redshift of $0.12$ and bulge-disk decomposition using
GIM2D \cite{gim2d} has now been completed for all 10 095 systems
(Allen et al. these proceedings). The survey has been used to constrain
estimates of the nearby luminosity function \cite{mgc1} as well as
measure the joint luminosity-surface brightness distribution
\cite{mgc4}. Plans are afoot to expand the MGC program with additional
$ugriz$ data from VST and $YJHK_s$ data from UKIRT/VISTA. The redshift
survey will be extended to $B=22$ mag with AAT/AA$\Omega$ and
eventually to $B=24$ mag with
Gemini/WFMOS. 

\section{The MGC-SMBH mass function}
\begin{figure}[h]
\begin{center}
\epsfig{figure=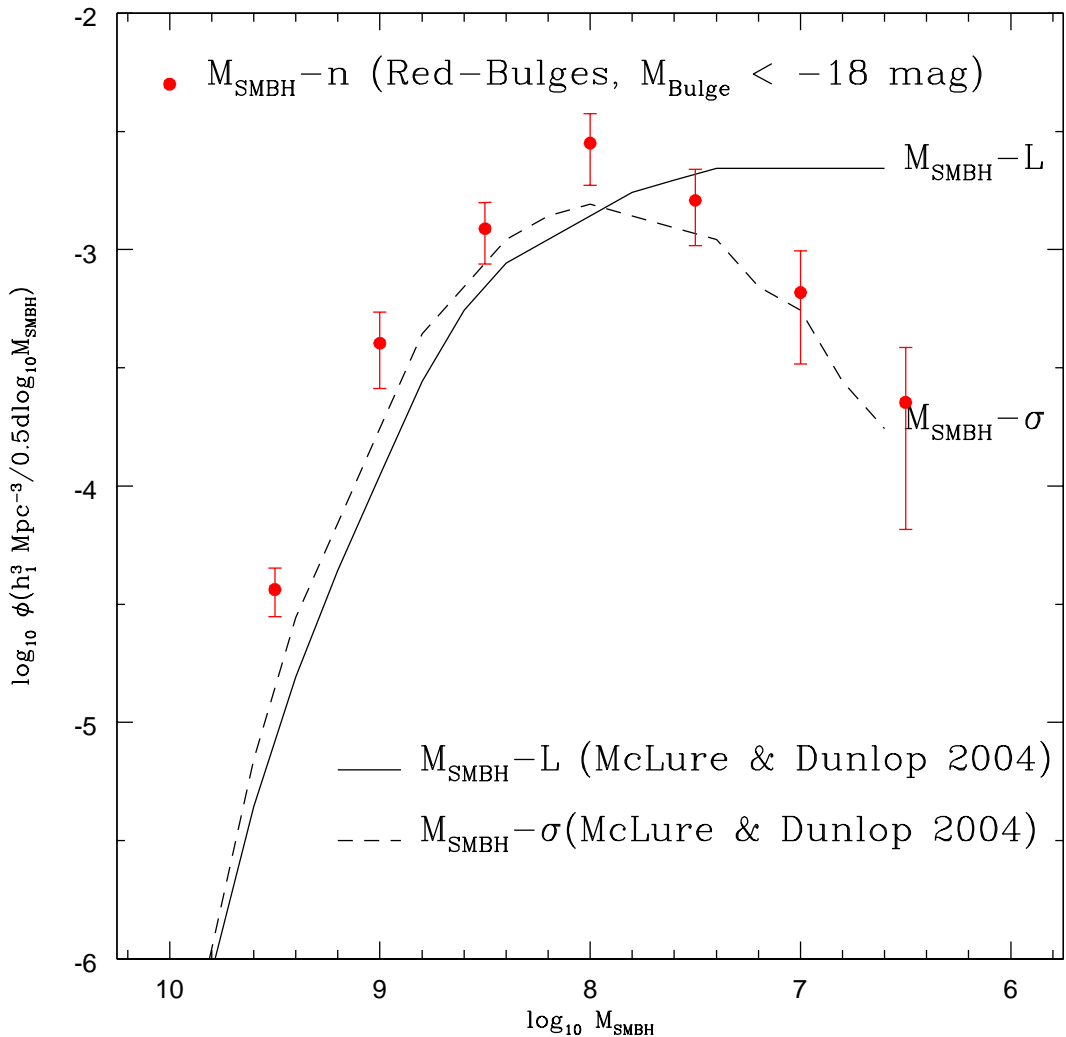,width=15.0cm}  
\end{center}
\caption{The two lines show the SMBH mass functions derived by McLure
\& Dunlop (2004) using either the $M_{BH}-\sigma$ ({\it dashed line})
or the $M_{BH}-L$ relation ({\it solid line}). Our results are show as data
points with errorbars and appear to follow the $M_{BH}-\sigma$ results
of McLure \& Dunlop reasonably closely. \label{fig:smbhmf}}
\end{figure}

To construct the SMBH mass function we need to combine the relation
shown in Eqn.~1 with the luminosity function of galaxy bulges (see
Liske et al., these proceedings). Our approach is to assign a black hole
mass to each galaxy (based on the S\'ersic-index) and then adopt a
weight for each galaxy dependent on the space-density of galaxies of
that luminosity divided by the number of galaxies contributing to that
calculation, i.e.,
\begin{equation}
W_i = \phi(M,\mu)/N(M,\mu) \equiv 1/V_{MAX}.
\end{equation}
Having established a list of weights, $W_i$, and masses, one can then
simply sum the product of these distributions, thus:
\begin{equation}
N(M_{BH}) \delta \log M_{BH} = \sum_i W(M_{BH_i})_i \delta \log M_{BH}.
\end{equation}
The result, shown in Fig.~\ref{fig:smbhmf}, includes genuine bulges
but excludes pseudo-bulges (distinguished by their blue colours, see
Liske et al. these proceedings).  Also shown are recent estimates by
\cite{McCD} based on $M_{BH}-\sigma$ (dashed line) and $M_{BH}-L$
(solid line) estimates for a sample of $\sim 9 000$ galaxies drawn
from the Sloan Digital Sky Survey. The integral of this value leads to
the total density of SMBHs which we tentatively find to be $\rho=3.4
\times 10^5$ M$_{\odot}h^3_{0.7}$Mpc$^{-3}$. This is somewhat higher
than the McLure \& Dunlop values most likely due to our inclusion of
later-type bulges.

\section{Conclusions} 
We have highlighted the correlation between the mass of the super
massive black hole and the S\'ersic index (i.e., the $M_{BH}-n$
relation). This is first presented in \cite{graham01} and followed up
in \cite{graham02}.
The S\'ersic index of spheroids and bulges is straightforward to
measure and we have now done so for 10 095 galaxies drawn from the
{\it Millennium Galaxy Catalogue}. Using the $M_{BH}-n$ relation
combined with our recent luminosity function estimate \cite{mgc4} we
have now constructed the nearby mass function of super massive black
holes. The shape of the distribution closely follows that found by
\cite{McCD} albeit with a slightly higher normalisation. We ascribe
this to our method including all bulges whereas \cite{McCD}
exclusively focused on spheroids and early-type spiral systems.  The
result presented here will be presented in a more robust fashion in
the forthcoming paper by Graham et al., in preparation.

\acknowledgements{
The Millennium Galaxy Catalogue consists of imaging data from the
Isaac Newton Telescope and spectroscopic data from the Anglo
Australian Telescope, the ANU 2.3m, the ESO New Technology Telescope,
the Telescopio Nazionale Galileo, and the Gemini Telescope. The survey
has been supported through grants from the Particle Physics and
Astronomy Research Council (UK) and the Australian Research Council
(AUS). The data and data products are publicly available from
http://www.eso.org/$\sim$jliske/mgc/ or on request from J. Liske or
S.P. Driver.}

\vfill 

\begin{thebibliography}{}{ 
\vspace{-0.1cm}\bibitem{conselice} Bershady M.A., Jangren A., Conselice C.J., 2000, AJ, 119, 2645
\vspace{-0.1cm}\bibitem{croton} Croton D., et al., 2005,  MNRAS, in press (astro-ph/0508046)
\vspace{-0.1cm}\bibitem{dv} de Vaucouleurs G., 1948, AnAp, 11, 247
\vspace{-0.1cm}\bibitem{mgc4} Driver S.P., et al., 2005, MNRAS, 360, 590
\vspace{-0.1cm}\bibitem{eke} Eke V.R., et al., 2005, MNRAS, 362, 1233
\vspace{-0.1cm}\bibitem{fj} Faber S.M., Jackson R.E., 1976, ApJ, 204, 668
\vspace{-0.1cm}\bibitem{ferrarese00} Ferrarese L., Merritt D., 2000, ApJL, 539, 9
\vspace{-0.1cm}\bibitem{ferrarese} Ferrarese L., Ford H., 2004, SSRv, 116, 523
\vspace{-0.1cm}\bibitem{kcf70} Freeman K.C., 1970, ApJ, 160, 811
\vspace{-0.1cm}\bibitem{gebhardt} Gebhardt K., et al., 2000, ApJL, 539, 13
\vspace{-0.1cm}\bibitem{ghez} Ghez A.M., et al., 2003, ApJ, 586, L127
\vspace{-0.1cm}\bibitem{graham01} Graham A.W., Erwin P., Caon N., Trujillo I., 2001, ApJL, 563, 11
\vspace{-0.1cm}\bibitem{graham01b} Graham A., Trujillo I., Caon N., 2001, AJ, 122, 1707
\vspace{-0.1cm}\bibitem{graham02} Graham A.W., et al., 2003, in Galaxy Evolution:
Theory and Observations, RevMex AA (SC), eds., V.Avila-Reese,
C.Firmani, C.S.Frenk, \& C.Allen, vol. 17, 196-197
\vspace{-0.1cm}\bibitem{gd05} Graham A.W., Driver S.P., 2005, PASA, 22, 118
\vspace{-0.1cm}\bibitem{mgc1} Liske J., et al., 2003, MNRAS, 344, 307
\vspace{-0.1cm}\bibitem{Matkovic} Matkovic A., Guzm\'an R., 2005, 362, 289
\vspace{-0.1cm}\bibitem{maciejewski} Maciejewski W., Binney J., 2001, MNRAS, 323, 831
\vspace{-0.1cm}\bibitem{McCD} McLure R.J., Dunlop J., 2004, MNRAS, 352, 1390
\vspace{-0.1cm}\bibitem{merritt} Merritt D., Ferrarese L., 2001, MNRAS, 320, L30
\vspace{-0.1cm}\bibitem{sersic63} S\'ersic J.,-L., 1963, BAAA, 6, 41
\vspace{-0.1cm}\bibitem{sersic68} S\'ersic J.,-L., 1968, Atlas de Galaxias Australes
(Cordoba: Observatorio Astronomico)
\vspace{-0.1cm}\bibitem{silkrees} Silk J., Rees M.J., 1998, A\&A, 331, 1
\vspace{-0.1cm}\bibitem{gim2d} Simard L., et al., 1999, ApJ, 519, 563
\vspace{-0.1cm}\bibitem{tremaine02} Tremaine S., et al. 2002, ApJ, 574, 740
\vspace{-0.1cm}\bibitem{trujillo} Trujillo I., Graham A.W., Caon N., 2001, MNRAS, 326, 869
\vspace{-0.1cm}\bibitem{valluri} Valluri M., Ferrarese L., Merritt D., Joseph C.L., 2005, in press (astro-ph/0502493)

} 


\end{thebibliography}
\end{document}